\documentclass[fleqn,twoside]{article}
\usepackage{espcrc2}

\usepackage{graphicx}
\usepackage[figuresright]{rotating}

\newcommand{\AmS}{{\protect\the\textfont2
  A\kern-.1667em\lower.5ex\hbox{M}\kern-.125emS}}
\newcommand{\half}{\textstyle{1\over2}}

\title{A transverse lattice QCD model for mesons\thanks{Presented by A. Patel.}}

\author{Apoorva D. Patel\address[CTS]{Centre for Theoretical Studies,
        Indian Institute of Science, Bangalore-560012, India}${}^,$%
        \address[SERC]{Supercomputer Education and Research Centre,
        Indian Institute of Science, Bangalore-560012, India}%
        and Raghunath Ratabole\addressmark[CTS]}
       
\begin{document}

\begin{abstract}
QCD is analysed with two light-front continuum dimensions and two
transverse lattice dimensions. In the limit of large number of colours
and strong transverse gauge coupling, the contributions of light-front
and transverse directions factorise in the dynamics, and the theory can
be analytically solved in a closed form. An integral equation is obtained,
describing the properties of mesons, which generalises the 't~Hooft equation
by including spin degrees of freedom. The meson spectrum, light-front
wavefunctions and form factors can be obtained by solving this equation
numerically. These results would be a good starting point to model QCD
observables which only weakly depend on transverse directions, e.g.
deep inelastic scattering structure functions.
\end{abstract}

\maketitle

One of us had pointed out that transverse lattice QCD (i.e. QCD with
two continuous dimensions on light-front and two transverse dimensions
on lattice), can be solved analytically in a closed form in the
$N\rightarrow\infty$ and strong transverse gauge coupling limits,
and the solution would possess many observed properties of hadronic
physics \cite{PATEL}.
Let us review the ``pros and cons'' of this approach,
before going on to some explicit results:

\noindent
$\bullet$ The $N\rightarrow\infty$ limit leaves out sea quarks,
still enough non-perturbative physics can be described in terms
of the remaining valence quarks.\\
$\bullet$ Anisotropic renormalisation group evolution can reach
the $g_\perp\rightarrow\infty$ limit, but it is outside the physical
weak coupling scaling region.\\
$\bullet$ Although lattice discretisation of transverse directions restricts
the available range of $p_\perp$, for any 2- or 3-point Green's function,
it is always possible to choose a reference frame such that $p_\perp=0$
for all the external states.\\
$\bullet$ Fermion doubling problem appears due to the transverse lattice,
but $N\rightarrow\infty$ and $(p_\perp)_{\rm ext}=0$ highly restrict
unwanted fermion modes.\\
$\bullet$ Linear confinement is built in to the theory, but the dynamics
leading to it is different in longitudinal and transverse directions.
In longitudinal direction, it is due to linear Coulomb force in (1+1)-dim;
while in transverse directions,
it is due to completely disordered gauge field.\\
$\bullet$ The gauge field can be exactly integrated out, and all meson
properties are described by a single gauge invariant integral equation
obeyed by their light-front wavefunctions. But this equation has to be
solved numerically to get physical results.\\
$\bullet$ For mesons, an infinite tower of states (radial excitations)
appears in each quantum number (spin-parity) sector,
while baryons require separate treatment as semi-classical solitons.\\
$\bullet$ The approach is tailor-made for the study of deep inelastic
hadronic physics. Experimental results for structure functions are
expressed in terms of light-front variables, and their scaling behaviour
implies that transverse dynamics does not contribute at the leading order.
So even with a coarse transverse lattice, the leading scaling behaviour
may remain largely undisturbed.

We study the limit of QCD defined by the action
($g$ is held fixed as $N\rightarrow\infty$):
\begin{eqnarray}
S &=& a_\perp^2 \sum_{x_\perp}\int d^2x~ \Big[ -{N \over 4g^2} \sum_{\mu\nu a}
      F_{\mu\nu}^a(x) F^{\mu\nu a}(x) \nonumber\\
  &+& \overline{\psi}(x) \Big( i\sum_\mu \gamma^\mu \partial_\mu
               - \sum_\mu \gamma^\mu A_\mu - m \Big) \psi(x) \nonumber\\
  &+& {\kappa \over 2a_\perp}\sum_n \Big\{
      \overline{\psi}(x) (r+i\gamma^n) U_n(x) \psi(x+\hat{n}a_\perp) \nonumber\\
  &+& \overline{\psi}(x+\hat{n}a_\perp) (r-i\gamma^n) U_n^\dag(x)
                                          \psi(x) \Big\} \Big] ~.
\end{eqnarray}
Here $\mu,\nu$ label the light-front directions, and $n$ labels the lattice
directions. Compared to a general transverse lattice QCD action,
we have dropped the terms,
${\rm Tr}\big[ D_\mu U_n(x) (D^\mu U_n(x+\hat{n}a_\perp))^\dag \big]$ and
${\rm Tr}\big[ U_n(x) U_m(x+\hat{n}a_\perp) U_n^{\dag}(x+\hat{m}a_\perp) U_m^{\dag}(x) \big]$,
because their coefficients vanish as $g_\perp\rightarrow\infty$.
The transverse lattice spacing, $a_\perp$, remains finite in this limit.
We have introduced the parameter $r$
in the nearest neighbour fermion hopping term;
$r=0$ corresponds to naive fermions (which can be exactly spin-diagonalised
to staggered fermions), while $r=1$ corresponds to Wilson fermions.
The anisotropy parameter $\kappa$ approaches unity in the continuum limit.
We use the Minkowski metric in light-front coordinates,
\begin{equation}
\half\{\gamma_\alpha,\gamma_\beta\} = g_{\alpha\beta} = \left(\matrix{
0 & 1 &  0 &  0 \cr
1 & 0 &  0 &  0 \cr
0 & 0 & -1 &  0 \cr
0 & 0 &  0 & -1 \cr}\right) ~.
\end{equation}

For this action,
the following functional integration steps can be performed exactly:\\
(i) As in the 't~Hooft model \cite{THOOFT}, the choice of gauge $A^+=0$
makes the functional integral over $A^-$ Gaussian.
Integrating out $A^-$ produces a linear potential between quark current
densities that is regulated using the principal value prescription.\\
(ii) The action is linear in the gauge links $U_n(x)$,
so they can be integrated out in the $N\rightarrow\infty$ limit.
The result is a hopping term between adjacent planes
for colour singlet local fermion bilinears.\\
(iii) Fermion fields can be eliminated in favour of bosonic meson fields,
$\sigma_{\alpha\beta}(x,y)\equiv\overline{\psi}_\alpha(x)\psi_\beta(y)$.\\
(iv) The bosonised action is explicitly proportional to $N$,
and so its stationary point value gives the generating functional for
connected Green's functions in the $N\rightarrow\infty$ limit.

In what follows, we describe the results for Wilson fermions.
In this case, the projection operator structure of the fermion
hopping term simplifies the formulae,
and the results are simple modifications of those for the 't~Hooft model.
Results for a general value of $r$,
including expressions for the 1PI effective action,
are available elsewhere \cite{RAGHU}.

For Wilson fermions, the transverse tadpole insertions renomalising
the quark mass vanish, and the quark propagator along the light-front
is the same as in the 't~Hooft model:
\begin{equation}
S(p) = {i \over p\!\!\!/ - m - \Sigma(p) + i\epsilon} ~,~~
\Sigma(p) = {-g^2 \gamma^+ \over 2\pi a_\perp^2 p^+} ~.
\end{equation}
Consequently, the chiral condensate is:
\begin{equation}
\langle\overline{\psi}\psi\rangle_{4d}
~=~ {2 \over a_\perp^2}\langle\overline{\psi}\psi\rangle_{2d}
~\mathop{\longrightarrow}\limits_{m\rightarrow0}~
-{N \over a_\perp^3\sqrt{3}}\left({g^2 \over \pi}\right)^{1/2} ,
\end{equation}
where the factor of $2$ arises from the change in dimensionality of spinors,
and we have used the 't~Hooft model result obtained using split-point
regularisation and operator formulation \cite{BURKARDT}.

The meson states satisfy a homogeneous Bethe-Salpeter equation.
Let the meson wavefunction with a spin-parity structure $\Gamma$ be,
\begin{equation}
\phi_\Gamma(p,q) = \langle \overline{\psi}(p-q) \Gamma \psi(q)
                 | {\rm Meson}_\Gamma (p) \rangle ~.
\end{equation}
There are two types of quark-antiquark interactions:
a gluon exchange in the longitudinal direction,
and bilinear fermion hopping in the transverse directions.
Without loss of generality, we choose the reference frame
where transverse momenta of all external states vanish. 
We also scale the ``$+$'' momentum components to the longitudinal
momentum fractions, by setting $p^+=1$.
Thus $p=(p^+=1,p^-=M^2/2,p_\perp=0)$.
The interactions then become independent of the ``$-$'' and ``$\perp$''
momentum components, and we can project the Bethe-Salpeter equation
on to the light-front:
\begin{equation}
\Phi_\Gamma(q^+ \equiv x)
= \int dq^-~\frac{d^2q_\perp}{(2\pi)^2}~\phi_\Gamma(p,q) ~.
\end{equation}

For a quark-antiquark pair of masses $m_1$ and $m_2$, we obtain
($\beta \equiv g^2/\pi a_\perp^2$, P$\equiv$principal value),
\begin{eqnarray}
\!\!\!\!\!\! &  & \!\!\!\!\!\!\!\!\!\! \mu^2(x)\Phi(x) \equiv \left[M^{2}-\frac{m_1^2-\beta}{x}-\frac{m_2^2-\beta}{1-x}\right]\Phi(x) \nonumber\\
\!\!\!\!\!\! &=& \frac{1}{2x(1-x)}\left[\frac{m_1^2}{2x}\gamma^+ +x\gamma^{-} +m_1\right] \nonumber\\
\!\!\!\!\!\! &\times& \int_0^1 \frac{dy}{2\pi}\Big\{ -\frac{g^2}{a_\perp^2}\mathrm{P}\frac{1}{(x-y)^2}\gamma^+\Phi(y)\gamma^+ \\
\!\!\!\!\!\! &  & \qquad\qquad +\kappa^2\Big[2\Phi(y) + \sum_{n}\gamma^n\Phi(y)\gamma^n\Big] \Big\} \nonumber\\
\!\!\!\!\!\! &\times& \left[\frac{\mu^2(x)}{2}\gamma^+ +\frac{m_2^2}{2(1-x)}\gamma^+ +(1-x)\gamma^- -m_2\right] .\nonumber
\end{eqnarray}
This is a matrix integral equation in Dirac space,
and physical meson states have to be obtained by diagonalising it.
We find it convenient to decompose the meson wavefunctions in the basis
that is a direct product of the usual Clifford algebra bases
in continuum and lattice directions, i.e.
$\Phi = \sum _{C,L}\Phi _{C;L}\Gamma ^{C;L}$,
where $\Gamma^{C;L}$ belongs to
$\left\{1,\gamma^+,\gamma^-,\half[\gamma^+,\gamma^-]\right\}
\otimes \left\{1,\gamma^1,\gamma^2,\half[\gamma^1,\gamma^2]\right\}$.
With this choice, the $16$ components of eq.(7) block-diagonalise in to
four sets of $4$ components each, labeled by the lattice index ``$L$''.

In general, eq.(7) has to be solved numerically.
But even without an explicit solution, its structural closeness
to the 't~Hooft equation allows us to infer the following meson properties,
based on the known results of the 't~Hooft model:\\
$\bullet$ The singular part of the interaction kernel is the same as in
the 't~Hooft model. It is this part which determines the behaviour of
finite norm solutions at the boundaries, $x=0$ and $x=1$ \cite{THOOFT},
and only the components $\Phi _{-;L}$ contribute to it. We thus expect
$\Phi _{-;L}$ to vanish at the boundaries as $x^{\beta _{1}}$ and
$(1-x)^{\beta _{2}}$, where
\begin{equation}
\pi\beta_i \cot(\pi\beta_i) = 1-\frac{m_i^2\pi a_\perp^2}{g^2} ~.
\end{equation}
$\bullet$ The non-singular part of the interaction kernel, arising from
the transverse lattice dynamics, is a wavefunction at the origin effect.
It provides a shift independent of the longitudinal momentum $x$,
but dependent on the spin-parity of the meson.\\
$\bullet$ The light-front wavefunctions are gauge invariant by construction.
Their restriction to the finite box, $x\in [0,1]$,
guarantees that the spectrum of $M^2$ is purely discrete \cite{COLEMAN}.
Meson states can thus be labeled by a radial excitation quantum number $n$,
in each spin-parity sector.\\
$\bullet$ For large $n$, the behaviour of eigenvalues and eigenfunctions
of the integral equation is governed only by the singular part of the
interaction kernel. Then as in the 't~Hooft model \cite{THOOFT},
\begin{equation}
\left.\begin{array}{c}
M_n^2 \simeq n\pi g^2/a_\perp^2 \\
\Phi_{-;L}^{(n)}(x)\simeq \sqrt{2}\sin (n\pi x)\end{array}\right\} ~n\gg 1 ~.
\end{equation}
The zero-point energy shift, provided by the non-singular part of the
interaction kernel, leads to parallel infinite towers of meson states
(in $M^2-n$ plane), labeled by spin-parity quantum numbers.\\
$\bullet$ The theory possesses exact, though not manifest, parity symmetry.
As a consequence, meson states in each tower should alternate in parity as
in the 't~Hooft model. Another consequence of parity symmetry is that
the equations for the components $\Phi _{C;n_{1}}$ and $\Phi _{C;n_{2}}$
are degenerate. So helicity$=\pm$ states (i.e. $n_1\pm in_2$ combinations)
have identical masses, e.g. in case of vector and axial mesons.
Explicit breaking of rotational symmetry makes the equations for the
helicity$=0$ state and the helicity$=\pm$ states non-degenerate.\\
$\bullet$ The transverse part of the interaction kernel vanishes
for the component $\Phi_{-;n_{1}n_{2}}$, for any value of $\kappa$.
It follows that $\Phi_{-;n_{1}n_{2}}$ is an eigenstate of the integral
equation with eigenvalue zero, when $m=0$.
This is the pseudoscalar Goldstone boson of the theory,
with the corresponding wavefunction $\Phi_{-;n_{1}n_{2}}^{(1)}=1$,
just as in the 't~Hooft model.
The fact that the chiral limit of the theory is at $m=0$ is a remarkable
result, given that Wilson fermions break chiral symmetry explicitly.\\
$\bullet$ Fitting the meson spectrum to experimental results
requires determination of the parameters $g^2$ and $\kappa$,
in addition to the quark masses.
It would be convenient to fix $g^2/a_\perp^2$ using the slope of
the infinite towers of states, and $\kappa$ by demanding
(almost) degeneracy of helicity$=0,\pm $ states.

We feel that with a simple structure and an exact chiral limit at $m=0$,
numerical investigations of eq.(7) will not be too complicated.
Detailed numerical analysis is in progress, to determine the meson
spectrum, structure functions, decay constants and form factors.
Other directions for explorations are the heavy quark expansion
(see e.g. \cite{HEAVYQ}), and solutions for baryons as solitons
of the bosonised 1PI effective action (see e.g. \cite{BARYON}).

\end{document}